\documentclass[%
 reprint,
 superscriptaddress,
showpacs,
nofootinbib,
 amsmath,amssymb,
 aps,
 prl,
floatfix,
]{revtex4-1}

\usepackage{color}
\usepackage{graphicx}% Include figure files
\usepackage{dcolumn}% Align table columns on decimal point
\usepackage{bm}% bold math
\begin{document}

\preprint{}

\title{Wake turbulence observed behind an upstream \\'extra' particle  in a complex (dusty) plasma}

\author{S. Zhdanov}
 \affiliation{Max Planck Institute for extraterrestrial Physics, PO Box 1312, Giessenbachstr., 85741 Garching, Germany}
\affiliation{Forschungsgruppe Komplexe Plasmen, Deutsches Zentrum f\"{u}r Luft- und Raumfahrt, Oberpfaffenhofen, Germany}%

\author{C.-R.Du}%
 \email{chengran.du@dhu.edu.cn}
\affiliation{College of Science, Donghua University, Shanghai 201620, People's Republic of China}

\author{M. Schwabe}
\affiliation{Forschungsgruppe Komplexe Plasmen, Deutsches Zentrum f\"{u}r Luft- und Raumfahrt, Oberpfaffenhofen, Germany}

\author{V. Nosenko}
\affiliation{Forschungsgruppe Komplexe Plasmen, Deutsches Zentrum f\"{u}r Luft- und Raumfahrt, Oberpfaffenhofen, Germany}

\author{H. M. Thomas}
\affiliation{Forschungsgruppe Komplexe Plasmen, Deutsches Zentrum f\"{u}r Luft- und Raumfahrt, Oberpfaffenhofen, Germany}

\author{G. E. Morfill}
 \affiliation{Max Planck Institute for extraterrestrial Physics, PO Box 1312, Giessenbachstr., 85741 Garching, Germany}
\affiliation{BMSTU Centre for Plasma Science and Technology, Moscow, Russia}

\date{\today}% It is always \today, today,

\begin{abstract}
An interaction of upstream extra particles with a
monolayer highly-ordered complex plasma is studied. A principally
new abnormal turbulent wake formed behind the supersonic upstream
particle is discovered. An anomalous type of the turbulence wake
clearly manifests in anomalously low thermal diffusivity and two
orders of magnitude larger particle kinetic temperature compared to
that of the 'normal' wake (Mach cone) observed by Du et al
[Europhys. Lett. \textbf{99}, 55001 (2012)].

\end{abstract}


\maketitle

\section{Introduction} Wake turbulence is known to form behind any object that is moving fast enough, for instance, behind an aircraft as it
passes through the air \cite{Liu2002}. This phenomenon is important
to study in many aspects, e.g., regarding the safety of the take-off of following flights or flights from crossing runways. A turbulent wake is
common to observe, but its particular structure depends strongly on
conditions at which the object passes through the medium. A famous
example is the Karman vortex street \cite{Karman1912} (see also,
e.g.,\cite{Durgin1971}). In particular, a well studied pattern of
this kind is formed behind the cylinder
\cite{Townsend1947,Henderson1997} or a flat plate
\cite{Ramaprian1982} in an air or liquid stream. Decaying grid
turbulence with self-similarity \cite{Tordella2012} adds more
spectacular multi-wake exemplars in fluids
\cite{Mohamed1990,Wilczek2008} and soap films \cite{Shak2007}.

\begin{figure*}
  \centering{
  \includegraphics[width=0.9\columnwidth]{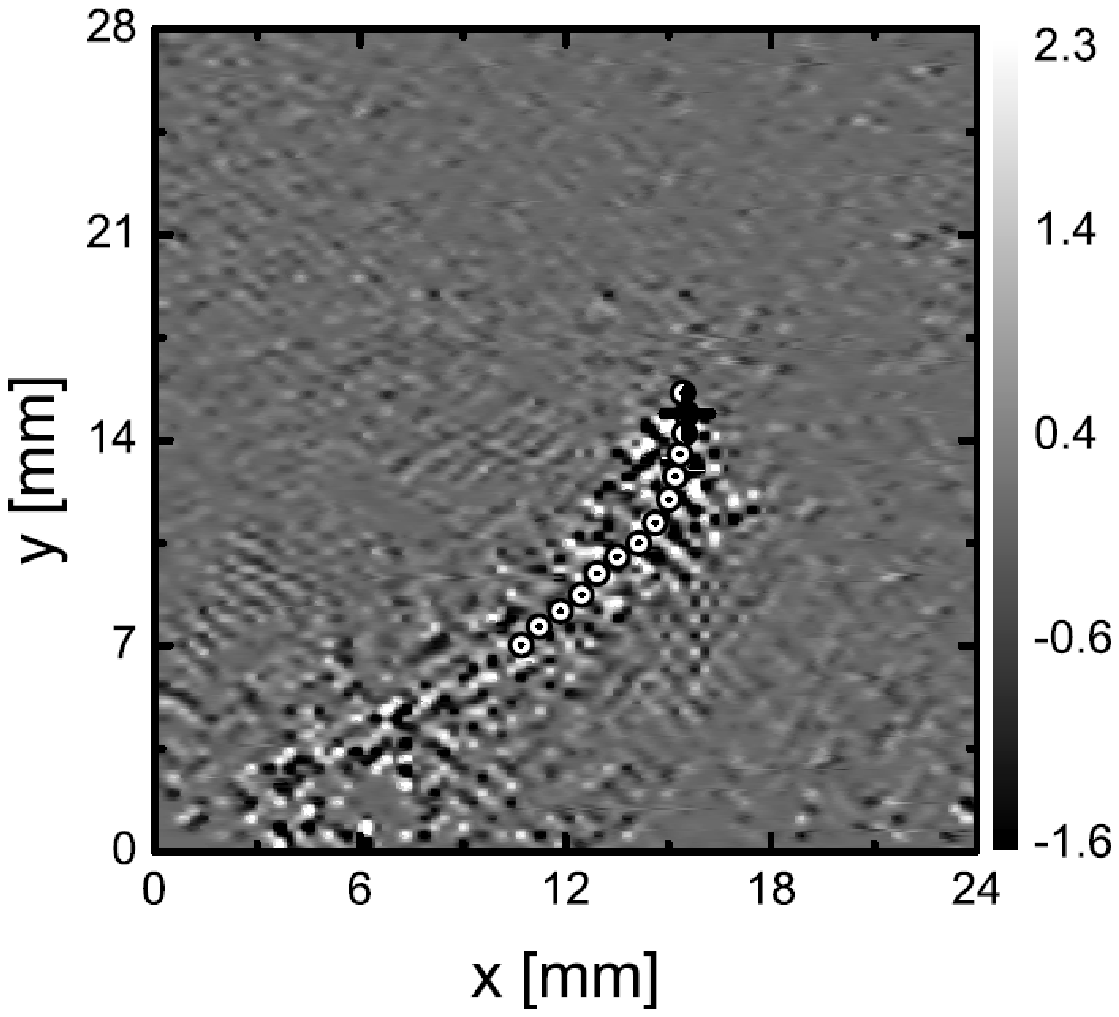}
  \includegraphics[width=0.95\columnwidth,viewport= 30 15 280 226]{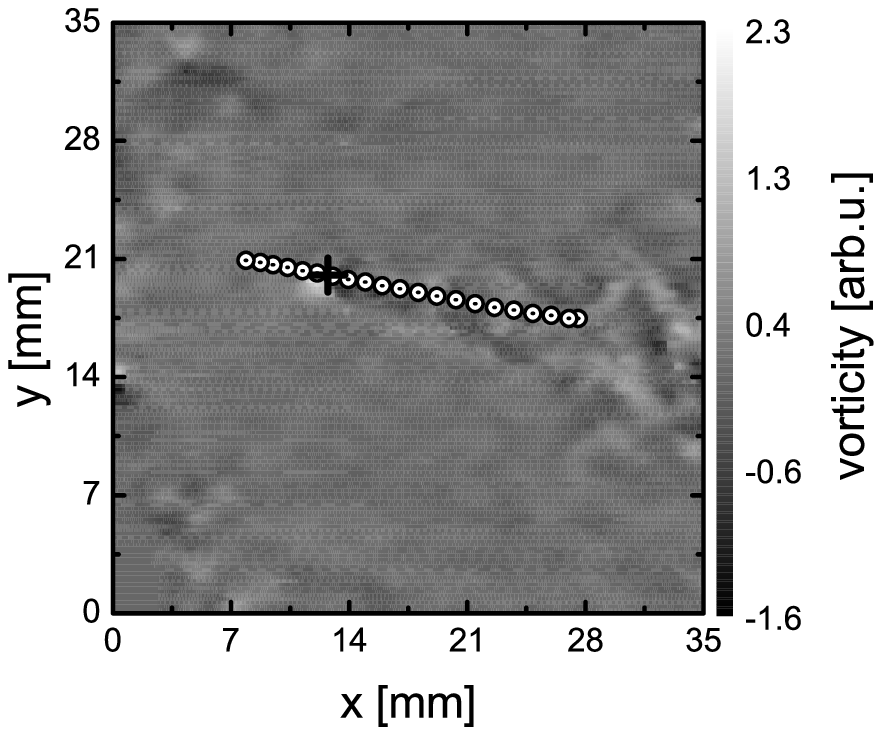}
  \includegraphics[width=0.9\columnwidth]{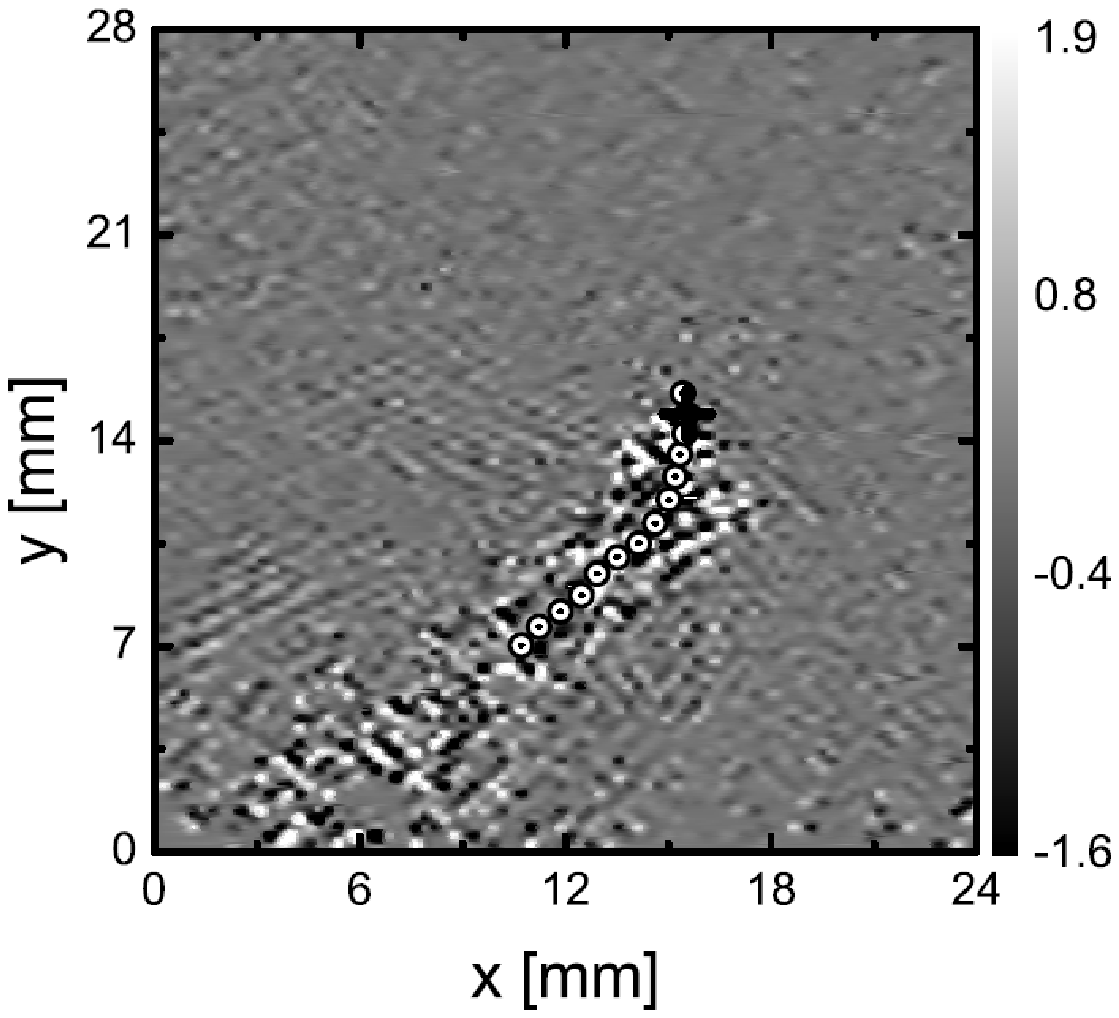}
  \includegraphics[width=0.95\columnwidth,viewport= 30 15 280 226]{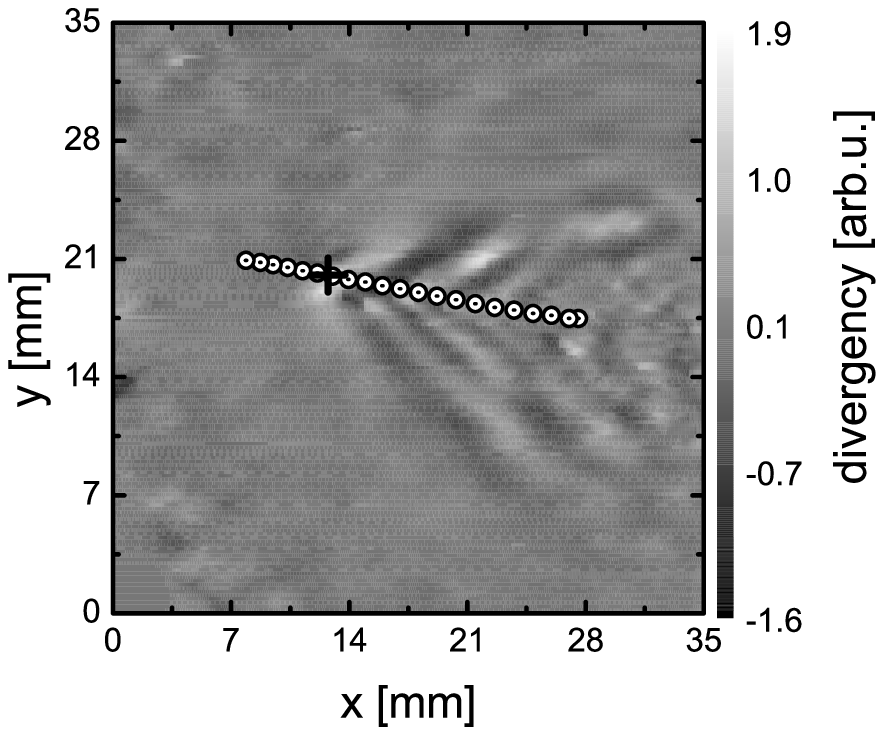}
  \includegraphics[width=0.9\columnwidth,bb= 19 15 290
  196]{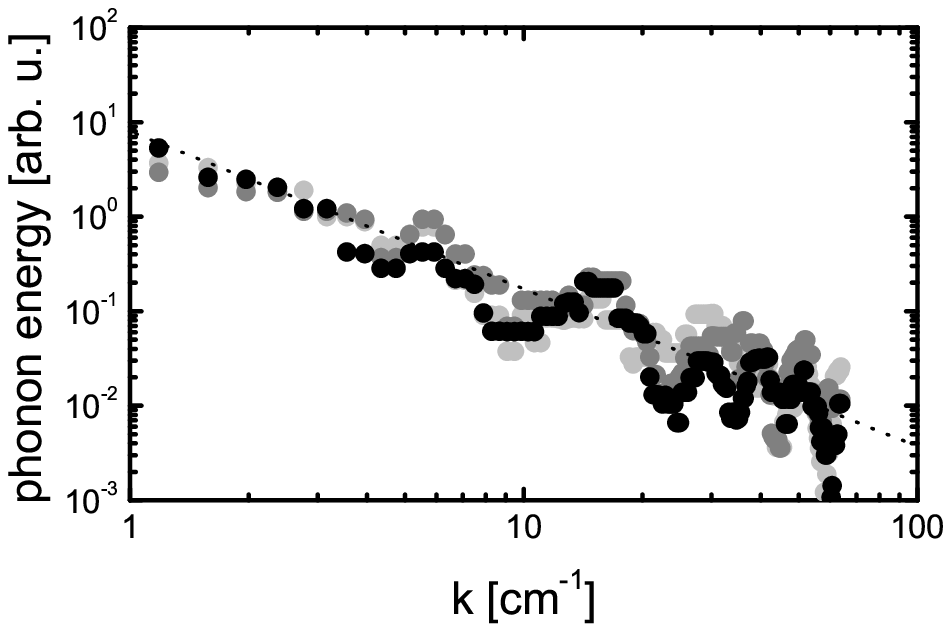}~~~
  \includegraphics[width=0.9\columnwidth,bb= 19 15 290
  196]{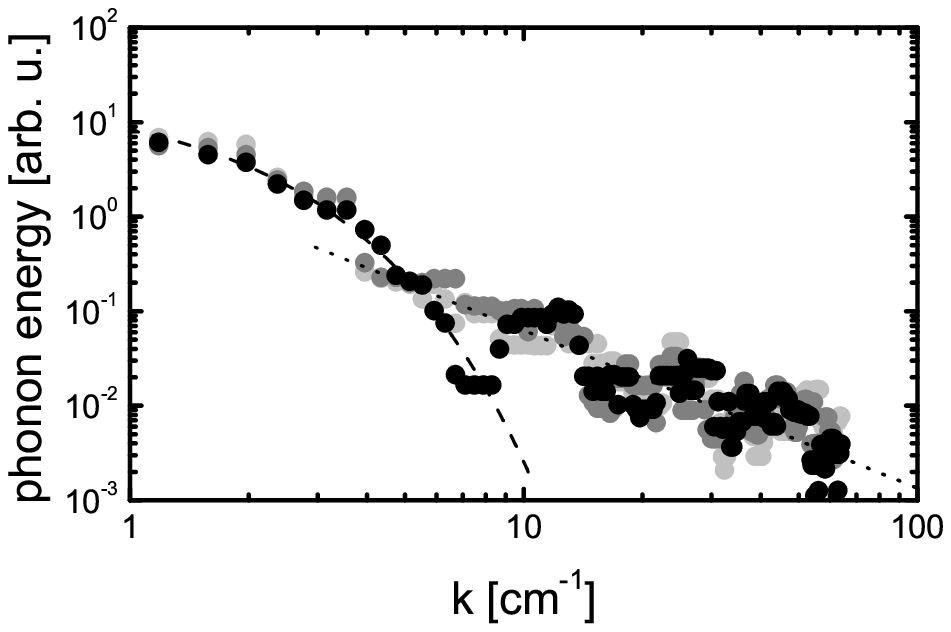}~~~~~}
  \caption{\label{fig:velocity} Vorticity (top panels), divergency (middle panels)
  maps of the velocity field, and phonon energy distributions (bottom panels) in
  the wake of the strongly scattered (left column) and the channeled (right
  column) upstream particle. The upstream particle tracks are shown by
  small open circles. The bent trace of the strongly scattered particle, which is moving upwards from the left lower corner to the right in the left maps, is immediately turbulent. In contrast, the very weakly turbulent wake behind the channeled particle is easy recognizable only at the cloud edge far away from the origin.  The channeled particle track is running from the right side to the left in the maps. In every map, the cross indicates the current position of the upstream particle at same time moment at which the maps were calculated. In the bottom panels, the black, grey, and light grey dots represent phonon spectra obtained for the time moment corresponding to the upper panels, and the two next moments delayed by 0.016~s respectively. The dotted line indicates the Kolmogorov turbulent spectrum $\propto k^{-5/3}$, and the dashed line the equilibrium phonon spectrum $\propto\exp\left(-k/k_T\right)$,
  $k_T=1.1\pm0.04$~cm$^{-1}$.}
\end{figure*}

The wake effect involving elements of turbulent pulsations is known
to be important when a sand flow interacts with an immobile
aluminium cylinder \cite{Chehata2003}. A peculiarly structured
Mach-cone-type wake (or 'shock') is formed behind a stainless-steel
rod inserted into and moving in a shallow, vibro-fluidized granular
layer consisting of bronze spheres \cite{Heil2004}. The shocks
formed were interpreted as an example of Cerenkov radiation
generated by an object traveling through a medium faster than the
wave phase velocity. The turbulent wake formation is interesting
also for explorations of the onset of turbulence
\cite{Grossmann2000}.

The turbulent wake behind small objects is especially interesting to
study as it has important implications in many physical and
biophysical applications as diverse as self-propelled stochastic
microdevices, turbulence around small organisms, "bacterial"
quasi-turbulence, in insect flight, and many others
\cite{Mitchell1985,Lazier1989,Cisneros2007,Golestanian2009,Wang2000,Tur2013}.
Regarding the wake formed behind small size objects there are of
interest either induced turbulence
\cite{Kajishima2002,Kajishima2004} or interactions with the
turbulent flow \cite{Cisse2013}. The microzones surrounding small
organisms as well as the turbulent flow at small scales are also
well-known challenging issues \cite{Mitchell1985,Lazier1989}.

A great advantage of turbulence studies using complex plasmas is
that the particles that transmit the interaction can be visualized
directly \cite{Zhdanov2010,Zhdanov2015}. For instance, vortices in
complex plasmas could be an ideal test bed for studying the onset of
turbulence and collective effects on the kinetic level
\cite{Schwabe2014}. Complex plasmas consist of micro\-meter-sized
particles embedded in a low temperature plasma. Under certain conditions they are highly ordered and form \emph{plasma crystals}
\cite{Ikezi:1986,Thomas:1996,Shukla2002a}.

A Mach cone (or wake) is formed when the monolayer plasma crystal is
perturbed by a supersonic charged particle-projectile
\cite{Dubin2000,Samsonov2000,Samsonov2005,Lenaic2012,Du2012,Du2014}.
Experimentally, the Mach cones (wakes) are often used as a robust
mean to test the plasma crystal for diagnostic purposes
\cite{Havnes2002,Schwabe2011} as well as to heat it
\cite{Nunomura2005}. A 'repulsion-dominated' Mach cone is created by fast 'extra'
particle moving \emph{below} the main particle monolayer
\cite{Samsonov2000,Samsonov2005}. Such 'downstream' particles are
heavier than the particles in the main layer and, therefore, their
tracks are only weakly perturbed by the presence of the monolayer.
The wake formed behind the downstream particle is fine structured in
all three directions, involving in-plane and vertical displacements
of the particles in a monolayer lattice as was recently
discovered \cite{Lenaic2012}.

An 'attraction-dominated' Mach cone is initiated when the particle
passes \emph{above} the monolayer \cite{Du2012,Du2014}. Those
'upstream' particles are lighter than the particles inside the main
layer, and, therefore, are easier for it to influence. It gives a
unique opportunity to observe rather delicate phenomena accompanying
the interaction of the plasma crystal with upstream particles, e.g.,
leading to constraining (\emph{channeling}) of the path of the upstream
particle \cite{Du2012}. The character of the upstream extra particle
motion depends on the local structure of the lattice layer. The path
of the well-channeled particle remains smooth and is followed by a
regular highly structured wake \cite{Du2012}.
%The channel
%distortion could cause strong scattering resulting in appearance of
%the short-living quasi-turbulent wake.

In this letter we report on a \emph{turbulent wake} formed behind
the upstream particle traveling above a highly-ordered lattice
layer. This principally new dynamical pattern is formed due to
strong upstream particle--lattice layer interactions. The dynamics of
the turbulent wake in complex plasmas is of great interest to study
as it is perhaps the simplest paradigm of stochastic or quasi-chaotic
billiards driven externally \cite{Sinai2000,Evans2001,Dieker2015}.
To our best knowledge persistent turbulent wakes in a highly ordered lattice layer have not been reported so
far. Implications for diagnostic purposes are also briefly
discussed.

\begin{table}%[htbp]
    \caption {Compressional and transverse sound speed $C_{L,T}$, projectile mean velocity $\langle V\rangle$ and velocity $V^{inst}$ at time of Fig.~\ref{fig:velocity}, the spatial damping increment $\kappa$, and the thermal
    diffusivity $\chi$ for the 'normal' channeling (case 1,
    Fig.~\ref{fig:velocity} right panels) and 'abnormal' strong
    scattering event (case 2, Fig.~\ref{fig:velocity} left panels).}
    %The
    %PS particles with diameter of $11.36~\mu$m and mass density of
    %1.05~g$/$cm$^3$ were suspended in a weakly ionized plasma of an
    %argon discharge at the pressure of 0.65 Pa and the discharge power
    %of 20 W.
    \label{tab:1}
    \begin{tabular}{cccccccc}
        \hline\noalign{\smallskip}
        case & $C_L/C_T$& $\langle V\rangle$& $V^{inst}$ & $\kappa$ & $\chi$  \\
        \noalign{\smallskip}\hline\noalign{\smallskip}
        & mm/s & mm/s & mm/s & mm$^{-1}$ & mm$^2/$s  \\
        \noalign{\smallskip}\hline\noalign{\smallskip}
        1& 27/6$^\alpha$ & 29 & 28 & 0.3$^\beta$ & 16$^\beta$  \\
        2& 18/5$^\gamma$ & 24 & 22 & 3.7$^\epsilon$ & 3.0$^\epsilon$ \\
        \noalign{\smallskip}\hline
    \end{tabular}\\
    $^\alpha$~measured in Ref.~\cite{Du2012}; $^\beta$~measured in
    Ref.~\cite{Du2014}; $^\gamma$~with accuracy $\pm15\%$; $^\epsilon$~with
    accuracy $\pm24\%$.
\end{table}

\section{Experiment particulars} 
The interaction of upstream extra particles with a 2D plasma crystal was studied by using a modified Gaseous Electronics Conference (GEC) rf reference cell \cite{Du2012,Du2014}. An argon plasma was sustained using a capacitively coupled rf discharge at 13.56~MHz and rf power at 20~W \cite{Du2012}. Monodisperse polystyrene (PS) particles were used to create 2D plasma crystals suspended above the bottom rf electrode. The PS particles have a diameter of $11.36\pm0.12 \mu$m and mass density of 1.05 g/cm$^3$. The gas pressure was maintained at about 0.65~Pa; the corresponding neutral gas damping rate was 0.91~s$^{-1}$\cite{Epstein1924}. After the particles were injected into the plasma, they formed a single-layer suspension with a size of 50-60~mm. In addition, some upstream extra particles that levitated above the main layer were present perturbing the layer.\footnote{The rigorous mechanism driving the particles debated in Ref. \cite{Du2012,Du2014} remains outstanding, though.}

\begin{figure}[t]
    \centering{\includegraphics[width=0.9\columnwidth,bb= 18 15 309
        226]{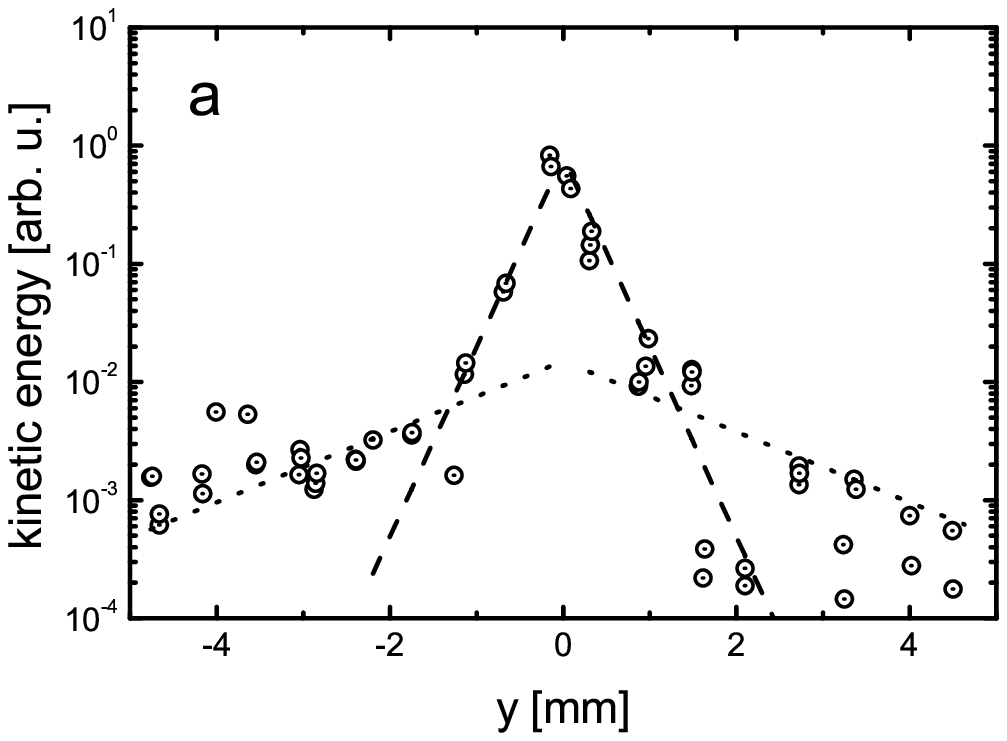}
        \includegraphics[width=0.9\columnwidth,bb= 18 15 309 226]{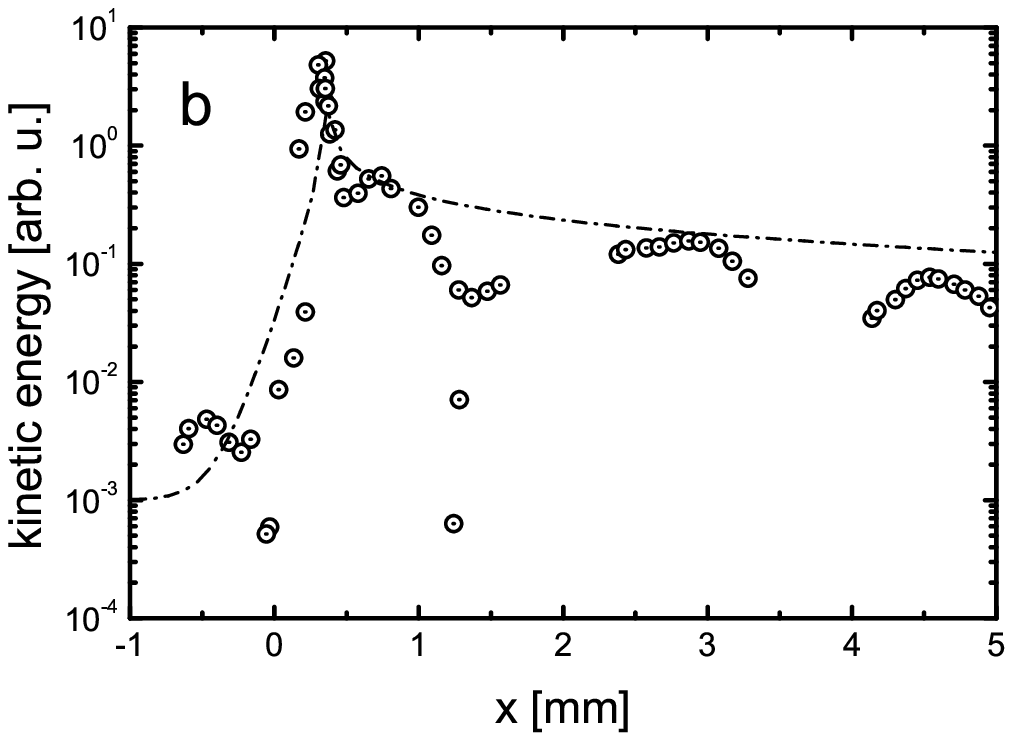}}
    \caption{\label{fig:ridges} Kinetic energy distributions inside the turbulent wake. The dots were obtained by imposing the particle positions reduced to the same origin ($\equiv$ current projectile position) taken in 10 consecutive frames inside narrow slabs of 0.2 mm width: (a) transversally to the projectile path at a distance 0.8 mm from the origin; (b) along the projectile path centered at the origin. The transverse distribution of the energy appeared to be nearly symmetric with respect to the origin (the projectile is located at $y=0$ in (a)). In the near field at $\mid y\mid\leq2$ mm the energy distribution is fitted well by the exponential function $E\propto\exp(-\kappa |y|)$ with the anomalously high inverse width $\kappa=L_{turb}^{-1}= 3.7\pm0.9 $~mm$^{-1}$ (the dashed line). In the far field the energy follows well the 'normal' exponential decay law with $\kappa =L_{norm}^{-1}= 16 $~mm$^{-1}$ (the dotted line). The dash-dotted line in (b) is the solution (\ref{eq:solution}-\ref{eq:kappa}) computed with the data of Table \ref{tab:1}.}
\end{figure}

The particle dynamics was recorded in high resolution using
a high-speed Photron camera at a frame rate of 250~fps. The recorded
frame sequences were then used to track the particle
positions from one frame to another as explained in detail in
Ref.~\cite{Du2012}. Before calculating the particle velocities, the obtained tracks were
filtered by applying a standard filtering
procedure to reduce the influence of industrial noise
($\sim 50$~Hz). Only the high quality filtered data were used further to compute
the particle velocity field and its differential
characteristics discussed below. The experimental parameters are
listed in Table~\ref{tab:1}.

\section{Laminar and turbulent wakes} Examples of laminar and turbulent wakes,
emerging behind a fast moving particle, are shown in Fig.~\ref{fig:velocity}. This
figure exhibits the vorticity and divergency maps of the velocity
fields formed behind the either strongly scattered
(Fig.~\ref{fig:velocity}, left panels) or channeled
(Fig.~\ref{fig:velocity}, right panels) upstream particle.
Strikingly different patterns are apparent there.

A peculiar wake with a collision-dominated quasi-chaotic pattern
inside its core is activated behind the strongly scattered particle
(Fig.~\ref{fig:velocity}, left panels). Curiously, the divergency
and the vorticity maps are difficult to distinguish. This indicates an approximate
equipartition of energy and momentum between
the shear and compression components of the turbulent velocity
field. Typically, the size of the wake core does not exceed about
2--3 mm; a relatively weak and more extended 'halo' is also present;
see Fig.~\ref{fig:ridges}.

In contrast, a far-fetched fine structure, patterning the wake, is
noticeable behind the channeled upstream particle
(Fig.~\ref{fig:velocity}, right panels). The structure of the
compressional wide-angle Mach cone represented by the divergency map
corresponds very well to theoretical expectations \cite{Dubin2000}.
The vorticity map helps to discover also a narrow-angle shear Mach
cone \cite{Nosenko2003} located mostly along the particle-projectile
path. The appearance of the shear Mach cone originates in a peculiar
character of deformation of the channel walls caused by the upstream
particle passing through the channel \cite{Du2012}. The half-open
angle $\theta$ of the cones in Fig.~\ref{fig:velocity} follows the
well-known Mach-cone-rule:
\begin{equation}
\sin\theta_{L,T}=C_{L,T}/v_d,
\end{equation}
where $C_{L,T}$ is the speed of sound, $v_d$ the projectile
velocity, and $L,T$ stands for longitudinal/transverse
waves. Note that simultaneous compressional and shear wakes excited by a
moving spot of laser light have been reported in \cite{Nosenko2003}. The neat 'coexistence'
of the naturally excited shear and compressional Mach cones is reported for the first time.

Turbulent wakes are as frequent as the laminar wakes discovered in
\cite{Du2014}, and likewise, their appearance cannot be controlled
experimentally \footnote{They appear regularly at large
scattering angles $\alpha$ when the kinetic energy transferred
$\Delta K\simeq K\alpha^2$ exceeds the wall barrier $\Delta K>W=\xi
Z^2e^2/a$, where $Z$ the particle charge, and $\xi$ the geometric
factor measured experimentally. For \cite{Du2012}, e.g., $\xi\simeq
0.4$.}

\begin{figure}[t]
    \centering{\includegraphics[width=0.9\columnwidth,bb= 18 15 309 226]{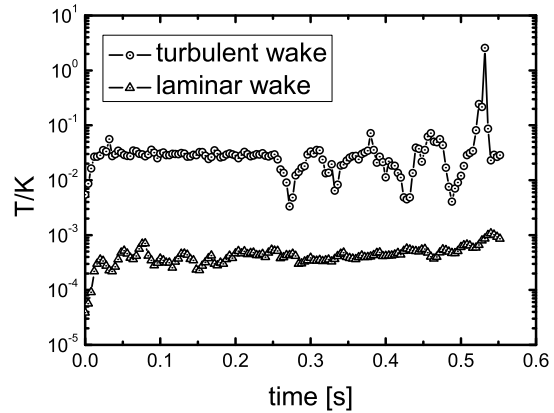}}
    \caption{\label{fig:chaos}  Kinetic temperature of the turbulent and laminar wake. The time is running from the moment when the upstream particle starts to encounter the monolayer. The kinetic temperature $T$ is reduced to the upstream particle kinetic energy $K$. To compare the near-field energies only the particles within a circle of radius 2~mm centered at the projectile were involved in calculations.}
\end{figure}

\begin{figure}[t]
  \centering{\includegraphics[width=0.88\columnwidth,bb= 18 15 273 220]{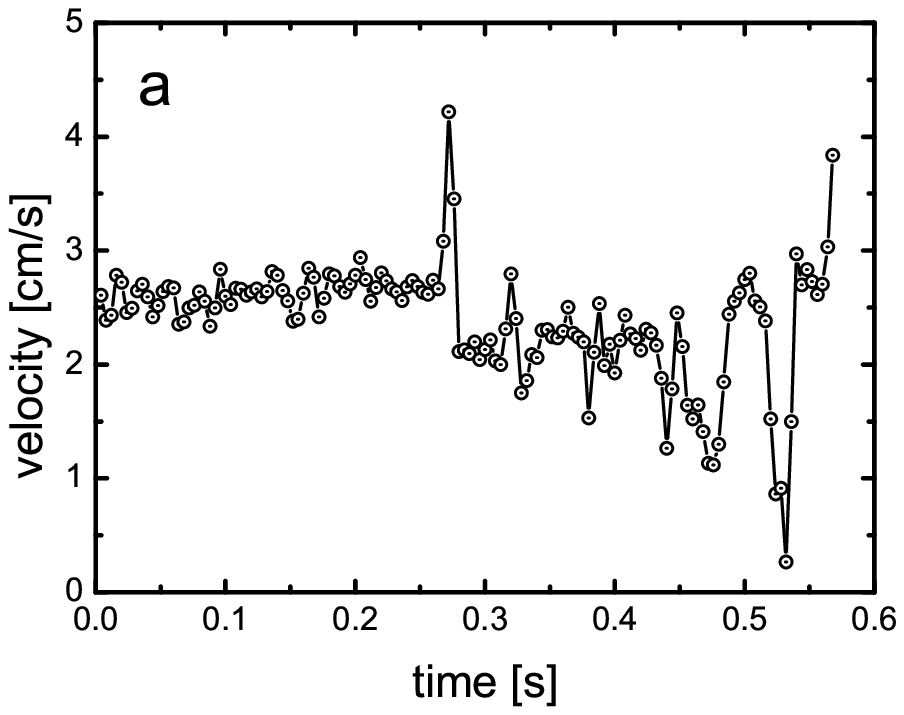}
  \includegraphics[width=0.9\columnwidth,bb= 18 15 286 220]{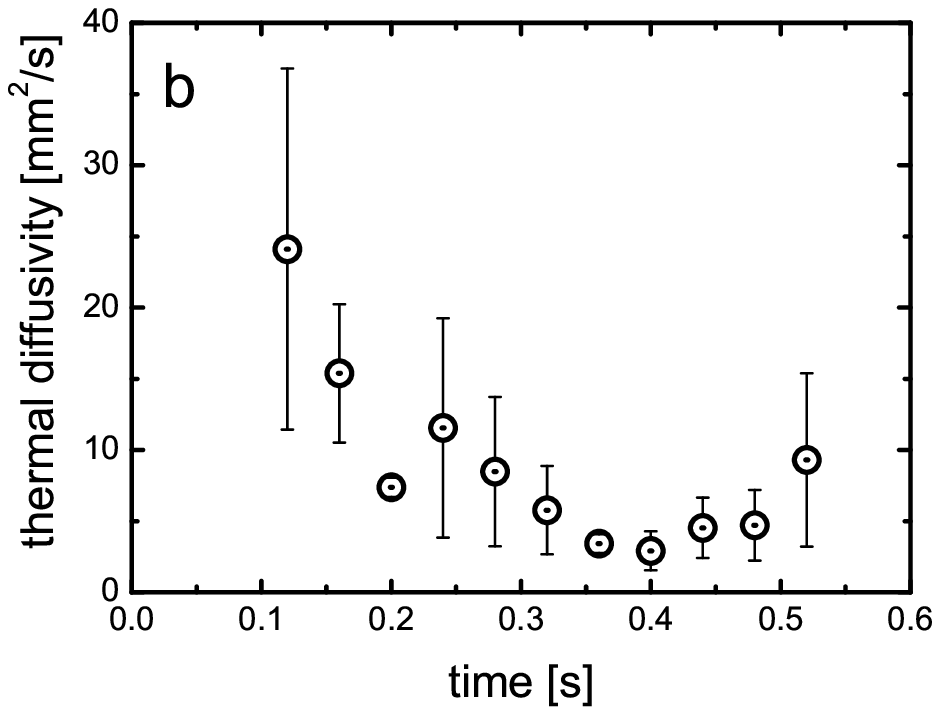}}
  \caption{\label{fig:result} Absolute velocity of the upstream particle (a) and the anomalous thermal diffusivity (b) obtained along the track of the upstream particle shown in Fig.~\ref{fig:velocity}, left panels. The particle enters the region above the monolayer at about $t=0$.
Strong collisions experienced by the upstream particle cause its velocity to significantly fluctuate along the path of travel. The thermal diffusivity is calculated using relationship (\ref{eq:kappa}). Note that the thermal diffusivity gradually increases towards the edge of the crystal.}
\end{figure}

To quantify the existence of turbulence in the wake we calculated
the isotropic part of the phonon energy of the crystal (see
Fig.~\ref{fig:velocity}, bottom panels)  In the case of channeled
upstream particle the main part of the energy corresponds to the
thermal phonon spectrum ($\propto\exp\left(-k/k_T\right)$
\cite{Matveev2010}) while in the case of strongly scattered upstream
particle almost all phonons exhibit the direct cascade power law
($\propto k^{-5/3}$) Kolmogorov \cite{Kolmogorov1941} spectrum. It
is worth noting also that the reduced kinetic temperature of the
turbulent wake is about two orders of magnitude higher than that of
the laminar Mach cone (see Fig.~\ref{fig:chaos}), and that its
thermal diffusivity is anomalously low (see Table~\ref{tab:1} and
Fig.~\ref{fig:result}).

\section{Diffusive wake model} Dynamically, the highly-collisional wake observed in
the given experiments is equivalent to a time-dependent billiard system,
and should be treated correspondingly \cite{Demers2015}. The
description of the wake in our case could be significantly
simplified, though. The energy of the wake particles changes
essentially at the time-scale of a single collision, the particles are
'forgetting' their initial velocities, and, hence, the wake energy
distribution evolves diffusively. In other words, the energy
distribution can be described by a Fokker-Planck equation, or, at
certain simplifying conditions, by a diffusion model
\cite{Bardos1997}.

The wake immediately behind the projectile blows up forming a
compact core which gradually transforms into an asymmetric halo. The
latter extends backwards slowly decaying along the path of travel;
see Fig.~\ref{fig:velocity} and Fig.~\ref{fig:ridges}. The
'core+halo' structuring persists while the core slowly
collapses when the projectile path approaches the monolayer center.
This can be interpreted as a weak large-scale variation of
the crystal transport coefficients. The heat transport can then be
modeled as follows:
\begin{equation}\label{eq:transport}
\partial_tT=-2\gamma T+\nabla\cdot\left[\chi\nabla T\right]+S(t,\mathbf{r}),
\end{equation}
where $T$ is the kinetic temperature of the monolayer particles, $t$
the time, $\gamma$ the friction drag coefficient \cite{Epstein1924},
$\chi$ the turbulent (anomalous) thermal diffusivity, $S$ the source
of perturbation, and $\mathbf{r}$ symbolizes the radius-vector. Let
us suppose, for example, a point-like perturbing source:
$S(t,\mathbf{r})\propto
\delta\left(x+Vt\right)\delta\left(y\right),$
%\begin{equation}\label{eq:source}
%S(t,\mathbf{r})\propto \delta\left(x+Vt\right)\delta\left(y\right),
%\end{equation}
where $x$ and $y$ are the coordinates in the monolayer plane,
$V\equiv \mathbf{V}_x$ the velocity of the projectile, and $\delta$
is the Dirac $\delta$-function. Assuming further a locally constant
thermal diffusivity $\chi$, one readily finds an analytical
solution:
\begin{equation}\label{eq:solution}
T(t,\mathbf{r})\propto
\exp\left(-\frac{V\eta}{2\chi}\right)K_0\left(\kappa R\right),~~
R=\sqrt{\eta^2+y^2},
\end{equation}
%\begin{eqnarray}\label{eq:solution}
%T(t,\mathbf{r})\propto \exp\left(-\frac{V\eta}{2\chi}\right)K_0\left(\kappa R\right),\nonumber\\
%R=\sqrt{\eta^2+y^2}, ~~\eta=x+Vt,
%\end{eqnarray}
where $\eta=x+Vt$, $K_0$ is the modified Bessel function, and
\begin{equation}\label{eq:kappa}
\kappa =\sqrt{\frac{2\gamma}{\chi}+\frac{V^2}{4\chi^2}}.
\end{equation}
The solution (\ref{eq:solution}-\ref{eq:kappa}) is comparatively
simple, approximates well the wave-form of the wake (see
Fig.~\ref{fig:ridges}) and, therefore, serves as a convenient
mean to measure the heat transport coefficient.

\section{Wake thermal diffusivity} It is important to mention that the projectile velocity $V$ and the damping rate $\gamma$ are independent of the heat transport model. Therefore, the only free parameter in the model
(\ref{eq:solution}-\ref{eq:kappa}) is the thermal diffusivity coefficient $\chi$. This fact can be conveniently used for diagnostic purposes.

The interparticle separation, which is approximately the mean free path of
the wake particles, is $a\approx 0.65$~mm, the wake core size is
$w=2-4$~mm, and the typical length of the projectile travel path $L=20-30$~mm in
the turbulent wake experiment. Therefore, the following inequalities
hold: $a<w<L$. Based on that we conclude that: (i) the diffusion wake model is applicable, at least qualitatively; (ii) since the wake size is much less than the length of the projectile path of travel, the thermal diffusivity coefficient can be treated as locally
constant.

Note that the energy distribution, being highly asymmetric along
the path of travel (\ref{eq:solution}) is fairly symmetric
transversally (cf. \cite{Du2014}). Moreover, the transverse
distribution is decaying approximately exponentially; see
Fig.~\ref{fig:ridges}a. Therefore, it is advantageous to explore the
transverse energy distribution, which significantly simplifies the fitting procedure. The thermal diffusivity
coefficient can be obtained by merging the measured spatial
damping increment of the experimental energy distributions to the
theoretical spatial damping increment (\ref{eq:kappa}). The results
of these calculations are shown in Fig.~\ref{fig:result}b. The spatial
resolution in this method is limited by the wake core size.

\section{Wake viscosity} The high collision rate of the particles inside the wake core makes it possible to suggest that turbulence has enough time to
become quasi-isotropic. This means, in particular, that the core
size must be of the order of the Kolmogorov turbulence scale-length
$L\simeq L_K\approx (\nu^3/\epsilon)^{1/4}$ \cite{Kolmogorov1941}, where $\nu$ is the wake viscosity, $\epsilon \approx \nu\Omega^2$ is
the (reduced) energy damping rate, and $\Omega$ is the vorticity.
The latter, in turn, could be roughly estimated as $|\Omega|\approx
V/L$. From relationship (\ref{eq:kappa}) it follows then that the
wake Prandtl number is:
\begin{equation}
P=\frac{\nu}{\chi}\simeq\frac{VL}{\chi}=\frac{2}{\sqrt{1+\xi}},~\xi=\frac{8\gamma\chi}{V^2}.
\end{equation}
As friction is weak $\xi\ll1$, the wake Prandtl number $P\simeq 2$.
To compare, it is certainly higher than the Prandtl number of the
argon gas in which the discharge was triggered \cite{Schwabe2011a}
still less than, e.g., that of water \cite{Landau1987}.

As expected, the anomalous wake appears to be more viscous than a
liquid complex plasma \cite{Nosenko2004}.

\section{Summary} To conclude, the interaction of an extra upstream particle
with a plasma crystal results in the appearance of an unusual
dynamic pattern which we termed the \emph{turbulent wake}. The
complex plasma of the turbulent wake exhibits the anomalously low
thermal diffusivity which makes it overheated compared to the normal
laminar wake (Mach cone). Experiments with turbulent wakes are a
promising mean to observe the anomalous transport in complex
plasmas.

\acknowledgments
  We acknowledge support from the European Research Council under the European Union's Seventh Framework Programme (FP7/2007- 2013)/ERC Grant Agreement No. 267499 and from the  National Natural Science Foundation of China (NNSFC), Grant No. 11405030. G.~Morfill wishes to acknowledge support from the Russian Science Foundation under grant 14-43-00053.

\end{document}